\documentclass{article}
\usepackage{array}
\usepackage[utf8]{inputenc}
\usepackage{graphicx}
\DeclareUnicodeCharacter{FB01}{fi}
\usepackage[utf8]{inputenc}
\usepackage{graphicx}
\title{Literature review}
\title{Fatigue Detection }

\author{Ashish Verma, Ankush Goyal , Davinderjit Kaur \\ \\ Computing Science\\ Multimedia\\University of Alberta\\ Edmonton, Alberta, Canada}

\begin{document}

\maketitle

\section{Introduction}
\par
Nowadays, there are many fatigue detection methods and majority of them  are  tracking eye in real time using one or two cameras to detect the physical responses in eyes. It is indicated that the responses in eyes have high relativity with driver fatigue. As part of this project, We will propose  a fatigue detection system based on pose estimation. Using pose estimation , We plan to mark the body joints in the upper body for shoulders and neck. Then , We plan to compare the location of the joints of  current posture with the ideal posture.
\section{Literature Review}
Abtahi et al. compares two different approaches to detect yawning[1]. The author uses two video datasets for the experiment. For the datasets, the participants are asked to sit in the driver's seat of a parked car. Then videos are captured for various mouth conditions such as resting, talking/singing, and yawning under different illumination conditions. The author keep exactly the same environmental parameters to collect videos for both datasets. The only change is the camera angle. For the first dataset, the camera is placed under the front mirror. In the second dataset, the camera is placed on the dash of the car. The author eventually describes that they achieved a higher degree of accuracy for yawning detection when the camera is placed on the dash board for the camera.[1]
\par
Martin et al. introduces the first large-scale dataset for recognizing driver activity. The data is collected in both manual and autonomous driving modes. A wide variety of information such as color, IR, depth and 3D body pose is collected from six views and categorized into 83 categories. The data is collected using five NIR- and the Kinect v2 cameras. The major challenges involves recognition of fine-grained behavior inside vehicle cabin, multi modal activity recognition and cross- view recognition benchmark. The author aims to bring the activity recognition models near to the real world application by highlighting and reducing the difficulting of dataset collection mechanism.[2]
\par
Szczapa et al. proposes an approach to deal with the problem of action recognition using body skeletons. As per the approach used, firstly, a trajectory is built on  Riemannian manifold from the body skeletons. Primary focus was to measure dynamic changes of the curves made of the landmark configurations. In order to smooth the curve and denoise the data points, a recent curve fitting algorithm is used. Instead of DTW (Dynamic Time Warping) which does not define a proper metric, Global Alignment Kernel (GAK) is used for temporal alignment which allowed derivation of valid positive – semidefinite  kernel when aligning to series. Finally, the generated kernel from GAK was used directly with Support Vector Machine (SVM) to classify actions.To check the validation of the approach different publicly available datasets are used like(a) KTH-Action dataset (61 frames) and (b)UAV-Gesture dataset (192 frames). In (a) the execution time is more because the GPU used is not powerful enough. Secondly, the low difference in computation time when switching from DTW to GAK. Also, using M2 reduces computational time by a factor of 2 as compared to M1.The execution time for treatment of skeletons in KTH-Action dataset=~0.499sec and for UAV-Gesture dataset=~0.614 sec. The proposed approach involved only body skeletons and the results are competitive with well recognised methods.[3]
\par
In 2014, IEEE  conference on computer vision and pattern recognition, a very interesting idea is  presented by Ravi Teja et. al. In this research, the authors enhanced the existing skeleton based human action recognition by new skeleton representation, that models 3d geometric relationship between various body parts. Since, the 3d representation lie on the euclidean group SE(3), authors presumes the representation lies in Lie group. Author uses various techniques for optimization and classification such as Dynamic Time warping, Fourier Temporal Representation. In the result, the author concludes that proposed approach outperformed numerous well recognized skeleton based recognition approaches.[4]
\par
Xiang Gao et al. in this research paper proposes the graph convolutional network over the recurrent neural network and convolutional neural network for irregular graph structure data. The author provides Spatio Temporal Modelling to enforce the sparsity of underlying graph. This is done for effective representation of consecutive frames. The optimized graph is fed into GCN with spectral graph convolutional.[5]
\par
In contrast to the above mentioned approach by Xiang where author represent GR-GCN, in this research paper Sijie Yan et. al. proposes spatial temporal graph convol.. network (STGCN). This approach is far more efficient for generalization capabilities and enhancing the power of expressions. In the evaluation matrix, author has leveraged more than 300000 clips from YouTube to cover approximately 400 action classes. In conclusion, the author claims to have improved performance in action recognition by using STGCN.[6]
\par
Bin Li et al. in this research paper author proposes Spatio-Temporal Graph Routing (STGR) scheme including two types of sub-networks SGR(Spatial Graph Router),  based on sub-group clustering along the spatial dimension this discovers the connectivity relationships among the joints and TGR(Temporal Graph Router), measure correlation degrees between temporal joint node trajectories. The
joint-connectivity graphs are then fed into ST-GCN in multiple routing ways.Then, introduces receptive field on graphs(coverage range that a node can draw information from).This technique is effective for learning high order connectivity relationships in regards to physically apart skeleton joints which further solve the weakness of predeﬁned human structure.[7]

Chao li et al. proposes a novel convo neural network, for both action classification and detection. In the paper, by combining the skeleton motion and skeleton transformation, they are able to achieve 83.2percent  and 89.3 percent  accuracy in partitioning schemes. This is done on 7 layer network with validation on a set of the NTU RGB+D dataset. Further on PKU-MMD dataset, they achieved 93.7 percent  mAP.[8]
\par

Pyry proposes a new technique based upon quantized trajectory snippets of tracked features.In it, feature trajectories(x and y directions) are produced for a number of features using KLT tracker. Further, these trajectories are transformed to trajectory snippets for each video. These snippets are quantized to a library of trajections. This approach slightly over performs the histogram of optical flow features in hollywood action dataset. Author uses standard KLT tracker to track approximately 100 fixed number of features and limit iterations to 20. When compared to existing motion features (optical flow, silhouettes, derivatives) trajections are able to take advantage of the positive features of each class: the computational efﬁciency of derivatives and sparse features, the performance of optical flow, and the deep structure of silhouettes.[9]
\par
Wentao Zhu et al proposes state of the art end to end connector deep long short term memory network for skeleton based action recognition. They designed an in depth dropout algorithm that not only operates on gates and cells but also output responses on the LSTM neurons. In this paper, the author experiments with different datasets. For instance, on the CMU dataset, they were able to achieve significant performance improvement based upon the proposed technique. [10]

\par
In this work[20],detection of human silhouettes in specific walking pose is done by  using spatio temporal templates.
Motion capture data is used to create sequences of 2-D silhouettes that are matched against short image sequences.
During training phase , statistical learning techniques are used to estimate and store the information about motion.
Templates covers six different camera views as well as seven different scales.
The silhouette matching to individual images is done using Chamfer distance which is computed using Distance Transform(DT) of canny edge images.
To avoid an exhaustive search for each silhouette in each frame, at each time step t,the silhouettes  of each template are searched in the image frames .
A lookup table is built for a fast access to the silhouettes detected in an image around a given location.
To avoid multiple responses for the same person, those detections are rejected that overlap with better ones.
Some body parts are clearly more informative than others. In this  specific case, since authors seek the pose where the legs are furthest apart, legs are clearly more important .
To handle this , the image is divided into several patches and different distributions are used for these patches.
The approach is done using only the walking pose but can be applied to any other motion.But creation of the respective motion database is required.[20]
\par
Detection-Based Online Multi-target Tracking via Adaptive Subspace Learning [19]
Multi target tracking can be divided into 3 key steps :
1) locate the multiple targets.
2) remember the identities of the targets.
3) predict the trajectories of the target.
Proposed method starts by formation of initial subspaces  by data then followed by data association .The final part of the approach is Adaptive subspace learning which  is used to 
track the change in appearance .Feasibility set is gathered which will contain a series of detections which may lead towards the trajectories.[19]

\par
EREL Selection using Morphological Relation.[18]
In this work , the aim is segmentation of arterial  wall boundaries from Intravascular Ultrasound (IVUS) images automatically.
Extremal Regions of Extremum Level (EREL)is a region detector that is capable of extracting repeatable regions from an image.
Each EREL is fitted by a closest ellipsoid since the intrinsic shape of coronary artery is similar to the shape.
Among these fitted ellipsoid , the one with the smallest difference between Hausdorff Distance (HD) and the ground truth is selected as segmentation
of luminal region.The proposed method in this paper uses the morphological characteristics of EREL regions as metrics to evaluate these ERELs and works on independent ERELs.[18]

\par
Statistical Relationship among Driver’s Drowsiness, Eye State and Head Posture. [17]
In this work , the authors aimed to show the statistical relationships among the driver's drowsiness ,eyestate and head posture.
In addition , multiple linear regression with ordinary least square method  and binary logistic regression was suggested for this.
The study suggests that the alert drivers normally look straight ahead whereas drowsy ones frequently look in other directions for a period of time.
The statistical relationships between the dependent variable and independent ones are interpreted from the regression coefficients which represent the quantitative effect
of each independent variable on the estimated change of the dependent one while other independent variables are kept unchanged. Regression coefficient is effectively obtained with
least squares (LS) methods. Ordinary least squares (OLS) method is used due to its simple computation.[17]

\par
A method of gaze direction estimation considering head posture.[16]
Head posture is analyzed based on the geometric features .Head postures are divided into 3 basic postures : Turn,  Nod , Swing.
Hough transform method is used for pupil detection .Hough transform can detect any curve with any known shape but shows exponential growth  along with growth of curve parameters.
Gaze direction is estimated using the combination of position of pupil in the normal state eye and head posture.[16]

\par
In A driver face monitoring system for fatigue and distraction detection [15] , an approach is introduced for driver fatigue detection and distraction based on the symptoms related to face and eye regions.
Head rotation(ROT) is considered as a symptom of distraction.
ROT is calculated using the difference between the position of head of the driver in frontal mode and current frame.
It is assumed that for the first 100 frames the head of the driver is in frontal position.
The features extracted from eye region are percent of eye closure (PERCLOS ) ,change in eyelid distance and eye closure rate(CLOSNO).
Face template matching is used for head rotation method .Spatio Temporal approach without explicit eye detection for feature extraction is used to extract facial features.
This method is based on change of the horizontal projection of the eye region during driving.
Normal values for the eye region features is estimated in the training phase  as different individuals have different eyelid behaviour.
The proposed fuzzy expert system processes four inputs and generates two outputs. The inputs are (1) PERCLOS, (2) ELDC, (3) CLOSNO, and (4) ROT, 
and outputs are (1) fatigue estimation and (2) distraction estimation. 
In order to build a fuzzy expert system, Mamdani fuzzy inference method ( min-max method) is applied on a set of fuzzy rules.[15]

\par
Predicting Student Seating Distribution Based on Social Affinity. [14]
In this paper , the authors propose  a method to learn the social affinity of the students.
The students’ seating distribution contains social data and can be used for analysing their social relationships.
The framework of  study is divided into 2 parts: constructing class 
social networks and predicting the average distribution of students’ seats.
For data collection , the instructor takes photos of the students at the beginning of the class.
Then the AdaBoost algorithm with skin-color model  is used to detect and recognize the students’ faces in the photo which taken in class.
Then the center projection principle and linear fitting algorithms are used for locate students’ position. And the intimate relationship between students can be determined by the Euclidean distance of their locations. Later, SAM is modified to build the eight neighbors around each student. In this way, the habit of how students choose companions are recorded. Finally, the prediction of the seating
distribution is obtained by long-term accumulation of the
statistics of seating distribution.[14]

\par
A Flexible Method for Time-of-Flight Camera Calibration Using Random Forest.[13]
A new method is proposed to calibrate the geometric distortion of ToF camera using random forest.
To form a depth pixel value in a ToF depth image, its corresponding IR
(Infrared) ray is projected onto the surface of the 3D object, with its re-
reflected light being detected by the sensor. By measuring the phase difference
between the radiated and reflected IR waves, the distance from the point on
object to the camera is thus calculated.
(1) the lens distortion caused by camera lens which distorts the direction of the IR ray. 
(2) the depth distortion error along the IR ray
Data is collected by placing a standard calibration chessboard in front of ToF camera in different positions and orientations 
Random forest is used to map high dimensional input space into a simple continuous space.Each decision tree is built using annotated set  of training samples.
Proposed method is adaptable and the calibration process uses only ToF camera and a standard chessboard.This method does not focus on the other noise sources of the ToF camera which have to be dealt independently.[13]

\par
A Survey on Vision-Based Hand Gesture Recognition.[12]
This paper compares the most common human-computer interaction products in recent years, which can be used to capture gesture data.
Authors have reviewed the latest technology and research findings and  compared the drawbacks and benefits of different methods of hand gesture recognition.
Static gesture recognition always use linear classifier and non-linear classifier, and the techniques of dynamic gesture recognition consists
of Time Delay Neural Network, Dynamic Time Warping, Finite State Machine, Hidden Markov Model (HMM) and the common methods Convolutional Neural Network (CNN).
Dynamic gesture recognition is the combination of these simple static gestures to convey the significance of their combination.
Researches on dynamic hand gestures are compared with techniques ,  number of gestures , Dataset used, limitations and average accuracy 
Various applications of vision based recognition are discussed such as Sign recognition , robot control and Game control.
The most commonly used gesture recognition method is a feedforward neural network named Convolutional Neural Network.But CNN needs a lot of pictures to train or reuse part of the neural network trained by massive data.Capsule Network can solve many problems of CNN,which makes it a possibility to replace CNN in the future  .[12]

\par
Affectional Ontology and Multimedia Dataset for Sentiment Analysis.[11]
In this paper , the ontological approach is compared with machine learning algorithms for sentiment analysis on social media from sparse and informal textual contents .
Affectional Ontology is proposed which represents a large number of emotional vocabulary. AFO can capture the sentiment from unstructured-informal messages .
The ontology development starts with knowledge and acquisition phase along with emotion words from psychological theories .Most of the previous datasets were domain specific ,so  domain-free sentiment multimedia dataset (DFSMD) is proposed with high quality messages and annotation labels.
Datasets was collected using Twitter APIs on all topics and independent emotions.Tweets  in English were used with inappropriate and non-useful tweets were discarded.
Dataset annotation was done using criteria set for both annotators and questions . only those Annotations were selected who spoke fluent english and were possessing a high level of education.
Every tweet was associated with sentiment question on a 3-level scale.
Tweets were divided into groups  for annotators to evaluate.
Constructed the DFSMD based on the following rules: the label was assigned to 
each tweet within the dataset based on a majority vote among annotators in each
of the 20 groups
Ontology sentiment analysis has 2 main steps :
1) Query the ontology
2) Calculate the sentence sentiment
For training machine learning classifiers , data is divided into two groups : 60percent for training and 40percent for testing.
Combining Bag-of-words features with lexicons' features used in the experiment boosted the performance 
In this work, the  considered sentiment affective states when building AFO  can extended to mood and emotion as well. [11]

\par
Human action recognition and prediction: A survey[21]
This is a survey paper to predict future actions.Though there are various deep-learning models to recognise human actions , there are some shallow methods which are still useful and works well on complex videos.Shallow methods such as improved dense trajectory with linear SVM has shown  promising results on large datasets.As different people drive differently and can have different postures, there are various intra - class variations which make algorithms difficult to label actions correctly and many state of the art algorithms are still failing in this case

\par
Research on Path Planning Method of an Unmanned Vehicle in Urban Road Environments[22]
In this paper, rather than considering unmanned vehicles as a particle, four wheel drive without side-slip is combined with mechanical constraints to find the path of vehicle.
Proposed approach – A compound algorithm based on A*(best for getting shortest path in static environments) is combined with Stochastic Fractal Search (SFC) which generates vehicle path that can be used for vehicle status information.Composite method is effective to resolve path planning in complex urban environments.

\par
Person Authentication by Air-Writing Using 3D Sensor and Time Order Stroke Context[23]
Rather than using a mouse or keyboard to capture person’s signature, in his paper, user’s signature are captured using a 3D sensor in air and then identification is verified using a novel reverse time-ordered shape context.
Backward representation can effectively filter out the signatures and simplifies the process of matching as a path finding problem. Then using a weighting scheme, path finding problem can be resolved  in real time through a dynamic time warping technique.
The proposed solution has significant accuracy rate of over 93.5 percent even without the requirement of initial gesturing.

\par 
Synthetic Vision Assisted Real-Time Runway Detection for Infrared Aerial Images[24]
Framework is designed for runway detection based on ROI based level method(proposed to speed up level set methods) and  synthetic vision which is a system of generating a rendered image of 3D scene topography and simulates what pilots actually saw in particular situation .Otsu’s thresholding method is generalised to trichotomize the region overlapping a virtual runway . An overview of generation of synthetic vision is given. The average processing time was reduced significantly thus allowing the model to be used for real time processing without any manual intervention of humans.

\par
Tactile Facial Action Units Toward Enriching Social Interactions for Individuals Who Are Blind[25]
The approach used in the paper focuses on facial action units because using the Facial Action Unit Coding System, any facial expression can be broken down into its fundamental building blocks called facial action units. These units provide a reliable and descriptive language for communication. In the paper, a user study is presented involving individuals who are blind, to evaluate the learnability, distinctness and naturalness of the proposed mapping. This study will be continued in future to analyse how feedback gathered during this research study can be used to further enhance the distinctness and naturalness of the proposed tactile facial action units.  

\par
A Deep Learning Approach to Predict Crowd Behavior Based on Emotion[26]
In this paper, approach to detect and predict crowd behaviors from spatio-temporal features based on emotion using a deep learning framework and multiclass Support Vector Machine (SVM) is proposed. Spatio-temporal descriptors using 3D Convolutional Neural Network (3DCNN) based on crowd emotions are extracted. The results are matched with 3 benchmark datasets :-
1.	Motion Emotion Dataset (MED)
2.	ViolentFlows
3.	UMN 
A 3D Convolutional deep learning framework is implemented to learn spatiotemporal features which can detect crowd emotions from a video and then a multiclass SVM is introduced to map the emotion features to 6 different crowd behaviors. 

\par
Driver fatigue detection system[27]
This paper presents a method for detecting the early signs of fatigue/drowsiness during driving. As a result of analysis in the paper ,the proposed system in paper will detect whether driver is able to drive or not. Various factors such as   heart rate variability (HRV), steering-wheel grip pressure, temperature difference between the inside and outside of the vehicle, make easy to determine  the driver’s fatigue level.
On the hardware side, the proposed system is made up of an analogical subsystem and other digital .Former does an adaptation of the signal to acquire it through an analogical to digital converter  while the later one filters and processes the resulting signal that it was gotten in the analogical phase.On the software side , system uses two types of software: one for the microcontroller ATmega128, and another one for the computer with a wireless link among both devices using Bluetooth.

\par
A system of driving fatigue detection based on machine vision and its application on smart device[28]
The proposed approach detects the face using front and deflected facial features.Then, candidate region of eye is determined according to the geometric distribution of facial organs .This system can be used in daily lives as it can be easy deployed in portable devices. The system is composed of four parts :-
1.     Image preprocessing
2.     Face detection,
3.     Eye state recognition
4.     Fatigue evaluation.
This paper proposed an improved strategy  to detect the driver fatigue based on machine vision and Adaboost algorithm.After detecting the face candidate region of eye is determined according to the geometric distribution of facial organs and then trained classifiers of open eyes and closed eyes are used to detect eyes in the candidate region quickly and accurately.

\par
Real time driver’s drowsiness detection system based on eye conditions[29]
This paper proposed an approach to detect driver’s drowsiness in real time. Under controlled conditions, result with 90 percent accuracy were achieved .Drowsiness of driver is detected using eye conditions .Using Viola Jones algorithm, eyes are detected through proposed crop Eye algorithm which segments the face in different segments in order to get left and right eye. The distance between eyebrow and eyelash is calculated and based on the threshold value , it is determined whether eyes are close or not. An alarm is triggered if eyes are found to be closed for consecutive five frames

\par
Real-time driver drowsiness detection for embedded systems using model compression of deep neural networks[30]
In this paper , an approach for detecting driver fatigue using deep learning methods which can be implemented on a low cost embedded board is proposed. There are publically accessible datasets for drowsiness detection. One is DROZY database , which contains multiple types of drowsiness-related data including signals such as EEG, EOG, ECG, EMG and near-infrared (NIR) images. The facial landmarks  are utilized as inputs to detect driver’s drowsiness and a compression technique of knowledge distillation is applied to be implemented on real-time embedded system For future, infrared cameras can be used to detect driver’s behaviour at night situation.

\section{References and Research papers}
1. Abtahi, Shabnam, Mona Omidyeganeh, Shervin Shirmohammadi, and Behnoosh Hariri. "YawDD: A yawning detection dataset." In Proceedings of the 5th ACM Multimedia Systems Conference, pp. 24-28. ACM, 2014.\\ \\
2. Martin, Manuel, Alina Roitberg, Monica Haurilet, Matthias Horne, Simon Reiß, Michael Voit, and Rainer Stiefelhagen. "Drive `|\&' Act: A Multi-modal Dataset for Fine-grained Driver Behavior Recognition in Autonomous Vehicles."\\ \\
3. Szczapa, Benjamin, Mohamed Daoudi, Stefano Berretti, Alberto Del Bimbo, Pietro Pala, and Estelle Massart. "Fitting, Comparison, and Alignment of Trajectories on Positive Semi-Definite Matrices with Application to Action Recognition." arXiv preprint arXiv:1908.00646 (2019).\\ \\
4. Vemulapalli, Raviteja, Felipe Arrate, and Rama Chellappa. "Human action recognition by representing 3d skeletons as points in a lie group." In Proceedings of the IEEE conference on computer vision and pattern recognition, pp. 588-595. 2014.\\ \\
5. Gao, Xiang, Wei Hu, Jiaxiang Tang, Jiaying Liu, and Zongming Guo. "Optimized skeleton-based action recognition via sparsified graph regression." In Proceedings of the 27th ACM International Conference on Multimedia, pp. 601-610. ACM, 2019.\\ \\
6. Yan, Sijie, Yuanjun Xiong, and Dahua Lin. "Spatial temporal graph convolutional networks for skeleton-based action recognition." In Thirty-Second AAAI Conference on Artificial Intelligence. 2018.\\ \\
7. Li, Bin, Xi Li, Zhongfei Zhang, and Fei Wu. "Spatio-Temporal Graph Routing for Skeleton-based Action Recognition." (2019).\\ \\
8. Li, Chao, Qiaoyong Zhong, Di Xie, and Shiliang Pu. "Skeleton-based action recognition with convolutional neural networks." In 2017 IEEE International Conference on Multimedia `|\&' Expo Workshops (ICMEW), pp. 597-600. IEEE, 2017.\\ \\
9. Matikainen, Pyry, Martial Hebert, and Rahul Sukthankar. "Trajections: Action recognition through the motion analysis of tracked features." In 2009 IEEE 12th international conference on computer vision workshops, ICCV workshops, pp. 514-521. IEEE, 2009.\\ \\
10. Zhu, Wentao, Cuiling Lan, Junliang Xing, Wenjun Zeng, Yanghao Li, Li Shen, and Xiaohui Xie. "Co-occurrence feature learning for skeleton based action recognition using regularized deep LSTM networks." In Thirtieth AAAI Conference on Artificial Intelligence. 2016.\\ \\
11.  Abaalkhail, Rana, et al. "Affectional Ontology and Multimedia Dataset for Sentiment Analysis." International Conference on Smart Multimedia. Springer, Cham, 2018.\\ \\
12.     Wang, Taiqian, Yande Li, Junfeng Hu, Aamir Khan, Li Liu, Caihong Li, Ammarah Hashmi, and Mengyuan Ran. "A Survey on Vision-Based Hand Gesture Recognition." In International Conference on Smart Multimedia, pp. 219-231. Springer, Cham, 2018.\\ \\
13. Xu, Chi, and Cheng Li. "A Flexible Method for Time-of-Flight Camera Calibration Using Random Forest." In International Conference on Smart Multimedia, pp. 207-218. Springer, Cham, 2018.\\ \\
14. Pei, Zhao, Miaomiao Pan, Kang Liao, Miao Ma, and Chengcai Leng. "Predicting Student Seating Distribution Based on Social Affinity." In International Conference on Smart Multimedia, pp. 29-38. Springer, Cham, 2018. \\ \\
15. Sigari, Mohamad-Hoseyn, Mahmood Fathy, and Mohsen Soryani. "A driver face monitoring system for fatigue and distraction detection." International journal of vehicular technology 2013 (2013). \\ \\
16.  Zhang, W. Z., Z. C. Wang, J. K. Xu, and X. Y. Cong. "A method of gaze direction estimation considering head posture." International Journal of Signal Processing, Image Processing and Pattern Recognition 6, no. 2 (2013): 103-111. \\ \\
17. Hien, Lam Thanh, and Thanh-Lam Nguyen. "Statistical Relationship among Driver’s Drowsiness, Eye State and Head Posture." Journal of Informatics and Mathematical Sciences 8, no. 1 (2016): 37-48. \\ \\
18.  Li, Yuying, and Mehdi Faraji. "EREL Selection using Morphological Relation." In International Conference on Smart Multimedia, pp. 437-447. Springer, Cham, 2018. \\ \\
19. Nigam, Jyoti, Krishan Sharma, and Renu M. Rameshan. "Detection-Based Online Multi-target Tracking via Adaptive Subspace Learning." In International Conference on Smart Multimedia, pp. 285-295. Springer, Cham, 2018. \\ \\
20.Dimitrijevic, Miodrag, Vincent Lepetit, and Pascal Fua. "Human body pose detection using Bayesian spatio-temporal templates." Computer vision and image understanding 104, no. 2-3 (2006): 127-139. \\ \\
21. Kong, Yu, and Yun Fu. "Human action recognition and prediction: A survey." arXiv preprint arXiv:1806.11230 (2018).
22. Ruixing, Yu, Zhu Bing, Cao Meng, Zhao Xiao, and Wang Jiawen. "Research on Path Planning Method of an Unmanned Vehicle in Urban Road Environments." In International Conference on Smart Multimedia, pp. 235-247. Springer, Cham, 2018. \\ \\
23. Chiu, Lee-Wen, Jun-Wei Hsieh, Chin-Rong Lai, Hui-Fen Chiang, Shyi-Chy Cheng, and Kuo-Chin Fan. "Person Authentication by Air-Writing Using 3D Sensor and Time Order Stroke Context." In International Conference on Smart Multimedia, pp. 260-273. Springer, Cham, 2018.\\ \\
24. Liu, Changjiang, Irene Cheng, and Anup Basu. "Synthetic Vision Assisted Real-Time Runway Detection for Infrared Aerial Images." In International Conference on Smart Multimedia, pp. 274-281. Springer, Cham, 2018.  \\ \\
25. McDaniel, Troy, Samjhana Devkota, Ramin Tadayon, Bryan Duarte, Bijan Fakhri, and Sethuraman Panchanathan. "Tactile Facial Action Units Toward Enriching Social Interactions for Individuals Who Are Blind." In International Conference on Smart Multimedia, pp. 3-14. Springer, Cham, 2018. \\ \\
26. Varghese, Elizabeth B., and Sabu M. Thampi. "A Deep Learning Approach to Predict Crowd Behavior Based on Emotion." In International Conference on Smart Multimedia, pp. 296-307. Springer, Cham, 2018 \\ \\
27. Rogado, Elena, José Luis Garcia, Rafael Barea, Luis Miguel Bergasa, and Elena López. "Driver fatigue detection system." In 2008 IEEE International Conference on Robotics and Biomimetics, pp. 1105-1110. IEEE, 2009 \\ \\
28. Kong, Wanzeng, Lingxiao Zhou, Yizhi Wang, Jianhai Zhang, Jianhui Liu, and Shenyong Gao. "A system of driving fatigue detection based on machine vision and its application on smart device." Journal of Sensors 2015 (2015)  \\ \\
29. Ullah, Asad, Sameed Ahmed, Lubna Siddiqui, and Nabiha Faisal. "Real time driver’s drowsiness detection system based on eye conditions." Int J Sci Eng Res 6, no. 3 (2015): 11-15 \\ \\
30. Reddy, Bhargava, Ye-Hoon Kim, Sojung Yun, Chanwon Seo, and Junik Jang. "Real-time driver drowsiness detection for embedded systems using model compression of deep neural networks." In Proceedings of the IEEE Conference on Computer Vision and Pattern Recognition Workshops, pp. 121-128. 2017 \\ \\
\end{document}